\documentclass{Pos}

\usepackage[english]{babel}
\usepackage{psfrag}	
\usepackage{cite,citesort}
\usepackage{amsfonts,amsmath,amssymb}
\newcommand{\eq}{ \; = \; }
\newcommand{\ye}{ \mathbf{Y}_e }
\newcommand{\ynu}{ \mathbf{Y}_{\nu} }
\newcommand{\bra}[1]{\left\langle #1 \right|}
\newcommand{\ket}[1]{\left| #1 \right\rangle}

\title{Revisiting CP-violation in Minimal Flavour Violation}

\ShortTitle{Revisiting CP-violation in MFV}

\author{Lorenzo Mercolli\\
        Center for Research and Education in Fundamental Physics,\\
 	Institute for Theoretical Physics, University of Bern, CH-3012 Bern, Switzerland\\
        E-mail: \email{lorenzo@itp.unibe.ch}}

\abstract{After a brief review of the minimal flavour violation hypothesis and its implementation in the MSSM, the most general parametrisation of the soft supersymmetry-breaking terms by means of the charged lepton and neutrino spurions is constructed. Thereby a type I seesaw mechanism is assumed to generate neutrino masses. This expansion introduces several new CP-violating phases, whose effects on the leptonic electric dipole moments are investigated.}

\FullConference{International Workshop on Effective Field Theories: from the pion to the upsilon \\
		February 2-6 2009\\
		Valencia, Spain}

\begin{document}

\section{Introduction}

The particular flavour structure of the Standard Model provides one of the major puzzles of particle physics. The excellent agreement of the SM predictions for hadronic observables, such as FCNC, $K$ and $B$ mixings etc., imposes very stringent constraints on the flavour structure of any extension of the SM. On the other hand, the discovery of neutrino masses and mixing confirmed the non-conservation of lepton flavour. It is not easy to argue that in a model for new physics the new particles carrying flavour quantum numbers should be above the TeV scale or that the whole SM extension should be flavour-blind. Minimal flavour violation (MFV) provides a model independent symmetry principle to accommodate a non-trivial flavour structure while allowing new physics below the TeV scale. This is achieved by implementing the same suppression mechanism for flavour transitions that is at work in the SM into the new physics lagrangian. Although MFV does not allow for new sources of flavour transitions, it introduces new CP-violating phases and their impact on flavour-diagonal CP-violating observables has to be analysed. 

Besides of being free of hadronic uncertainties, the lepton sector provides tight experimental limits for lepton flavour violating (LFV) processes and electric dipole moments (EDM). If the forthcoming experiments see any deviation from the SM prediction, which is essentially zero, it would be a clear signal of new physics. It is thus natural to study the effects of the new CP-violating phases on the leptonic EDMs. This is done within the MSSM with a type I seesaw mechanism to account for the neutrino masses.

\section{The principle of MFV}

First, let us review the MFV hypothesis starting from the flavour structure of the SM. It was pointed out in \cite{Chivukula1987} that the gauge sector of the SM is invariant under the flavour symmetry group 
\begin{equation}
 G_f \eq U(3)^5 \eq U(3)_Q \times U(3)_U \times U(3)_D \times U(3)_L \times U(3)_E  \;,
\end{equation}
i.e. we can rotate the fermion fields in family space, leaving the kinetic terms invariant. This flavour symmetry is extended by an additional $U(3)_N$ factor when three heavy right-handed neutrinos are added to the SM particle content. Of course this symmetry is broken by the Yukawa couplings, e.g. $E^I \ye^{IJ} L^I H_d$, where $I,J=1,2,3$ are family indices.

We can nevertheless make the whole SM lagrangian formally $G_f$ invariant by promoting the Yukawa matrices to spurions. This means that $\ye$ is required to transform  as 
\begin{equation}
 \ye \, \stackrel{G_f}{\longrightarrow} \, g_E \ye \, g_L^\dag \quad \mbox{with} \quad g_E \in U(3)_E \,, \; g_L \in U(3)_L \,,
\end{equation}
and analogously for $\mathbf{Y}_u$ and $\mathbf{Y}_d$. At the end of the day the spurions are fixed to the physical value of the couplings and the flavour symmetry is explicitly broken.
This formal invariance allows us to formulate the MFV hypotesis: \emph{The flavour symmetry $G_f$ is broken minimally} \cite{MFV}. This means that no matter what the origin of the flavour structure is, only the known sources, i.e. the SM spurions, are responsible for the breaking of the flavour symmetry. Requiring a model to respect MFV allows one to parametrise every flavour structure in terms of the SM spurions and thus the model will automatically inherit the hierarchical structure of the Yukawa couplings. We now have a symmetry principle at hand that can accommodate a non-trivial flavour structure. 

To start using MFV, we first have to clarify the spurion content in the SM. It is found \cite{Cirigliano2005} that assuming a type I seesaw mechanism \cite{Seesaw}, there are two spurions in the lepton sector: $\ye$, which has already been mentioned above, and $\ynu$. The background value for $\ye$ is given by the mass matrix of the charged leptons 
\begin{equation}
\ye  = \frac{1}{v_d} \, \left( \begin{array}{ccc} m_e & 0 & 0 \\ 0 & m_\mu & 0 \\ 0 & 0 & m_\tau \end{array} \right) \quad \mbox{with} \quad v_{u,d}=\bra{0} H^0_{u,d} \ket{0} \,. 
\end{equation}
In view of applying MFV to the MSSM, we already define the spurions with two Higgs doublets. Since the neutrino mixing parameters are not directly linked to the Yukawa coupling matrix $\ynu$, there is no simple relation to fix the spurion unambiguously. Assuming the heavy right-handed neutrinos to be degenerate, we get the expression \cite{Ynu}
\begin{equation} \label{eqn ynu}
 \ynu^\dag \ynu \eq \frac{M_R}{v_u^2} \, U \sqrt{\mathbf{m}_\nu} \, e^{2i \Phi } \sqrt{\mathbf{m}_\nu} \, U^\dag \quad \mbox{with} \quad  \ynu^\dag \ynu \, \stackrel{G_f}{\rightarrow} \, g_L \, \ynu^\dag \ynu \, g_L^\dag \;.
\end{equation}
$M_R$ is the seesaw scale, $\mathbf{m}_\nu= \mbox{diag}(m_{\nu1},m_{\nu2},m_{\nu3})$ the mass matrix of the light neutrinos and $U=U_{PMNS} \cdot \mbox{diag}(1,e^{i \alpha /2},e^{i \beta /2})$ is the Pontecorvo-Maki-Nakagawa-Sakata mixing matrix \cite{PMNS} with the two Majorana phases $\alpha$ and $\beta$. The three real parameters of the antisymmetric matrix $\Phi_{ij}=\epsilon_{ijk}\phi^k$ collect the arbitrariness mentioned above and provide an additional souce of CP-violation.

\section{MFV in the MSSM}

Knowing the two spurions relevant to the lepton sector, we can now implement MFV in the MSSM. In constrast to the SM, the flavour symmetry of the gauge sector is not only broken by the Yukawa couplings in the superpotential, but also by the soft supersymmetry-breaking terms
\begin{equation}
 \mathcal{L}_{soft} = - \tilde{L}^\dag \, \mathbf{m}_L^2 \, \tilde{L} - \tilde{E} \, \mathbf{m}_E^2 \, \tilde{E}^\dag + \tilde{E} \, \mathbf{A}_e \, ( \tilde{L} H_d ) + \dots \;,
\end{equation}
since the superpartners transform in the same way under $G_f$ as their corresponding SM fields. The matrices $\mathbf{m}_L^2$, $\mathbf{m}_E^2$, and $ \mathbf{A}_e$ ($3\times3$ matrices in family space) are not fixed by the theory itself and can cause huge flavour changes. The MFV hypothesis requires the soft-breaking terms to be formaly invariant under $G_f$ which means that we have to impose the following transformation properties
\begin{equation} \label{eqn softtrans}
 \mathbf{m}_L^2 \stackrel{G_f}{\longrightarrow} g_L \, \mathbf{m}_L^2 \, g_L^\dag \;, \quad \mathbf{m}_E^2 \stackrel{G_f}{\longrightarrow} g_E \, \mathbf{m}_E^2 \, g_E^\dag \;, \quad \mbox{and} \quad \mathbf{A}_e \stackrel{G_f}{\longrightarrow} g_E \, \mathbf{A}_e \, g_L^\dag \;.
\end{equation}
We can now expand these matrices in terms of the spurions $\ye$ and $\ynu$ \cite{MFV}. Remembering that 
\begin{equation}
 \ye^\dag \ye \, \stackrel{G_f}{\rightarrow} \, g_L \, \ye^\dag \ye \, g_L^\dag \quad \mbox{and} \quad \ynu^\dag \ynu \, \stackrel{G_f}{\rightarrow} \, g_L \, \ynu^\dag \ynu \, g_L^\dag \;,
\end{equation}
we can write the soft-breaking terms as sums of arbitrary powers of $\mathbf{A}\equiv\mathbf{Y}_{e}^{\dagger}\mathbf{Y}_{e}$ and $\mathbf{B} \equiv\mathbf{Y}_{\nu}^{\dagger}\mathbf{Y}_{\nu}$ in such a way that the correct transformation properties (\ref{eqn softtrans}) are reproduced. With the help of the Cayley-Hamilton identity it is then possible to reduce these sums to a finite number of terms \cite{Nikolidakis2007,Mercolli2009,Colangelo2008} and the final result reads
\begin{equation} \label{eqn MFVexp}
\begin{split}
\mathbf{m}_{L}^{2}  &  =m_{0}^{2}(a_{1}\mathbf{1}+a_{2}\mathbf{A}%
+a_{3}\mathbf{B}+a_{4}\mathbf{B}^{2}+a_{5}\{\mathbf{A},\mathbf{B}%
\}+a_{6}\mathbf{BAB}\\
&  \;\;\;\;\;\;\;\;\;\;\;+ib_{1}[\mathbf{A},\mathbf{B}]+ib_{2}[\mathbf{A}%
,\mathbf{B}^{2}]+ib_{3}(\mathbf{B}^{2}\mathbf{AB}-\mathbf{BAB}^{2}))\;,\\[2ex]
\mathbf{m}_{E}^{2}  &  =m_{0}^{2}(a_{7}\mathbf{1}+\mathbf{Y}_{e}%
(a_{8}\mathbf{1}+a_{9}\mathbf{B}+a_{10}\mathbf{B}^{2}+a_{11}\{\mathbf{A}%
,\mathbf{B}\}+a_{12}\mathbf{BAB}\\
&  \;\;\;\;\;\;\;\;\;\;\;+ib_{4}[\mathbf{A},\mathbf{B}]+ib_{5}[\mathbf{A}%
,\mathbf{B}^{2}]+ib_{6}(\mathbf{B}^{2}\mathbf{AB}-\mathbf{BAB}^{2}%
))\mathbf{Y}_{e}^{\dagger})\;,\\[2ex]
\mathbf{A}_{e}  &  =A_{0}\mathbf{Y}_{e}(c_{1}\mathbf{1}+c_{2}\mathbf{A}%
+c_{3}\mathbf{B}+c_{4}\mathbf{B}^{2}+c_{5}\{\mathbf{A},\mathbf{B}%
\}+c_{6}\mathbf{BAB}\\
&  \;\;\;\;\;\;\;\;\;\;\;\;\;\;+d_{1}i[\mathbf{A},\mathbf{B}]+d_{2}%
i[\mathbf{A},\mathbf{B}^{2}]+c_{7}i(\mathbf{B}^{2}\mathbf{AB}-\mathbf{BAB}%
^{2}))\;,
\end{split} 
\end{equation}
where $m_0$ and $A_0$ set the supersymmetry-breaking scale. So far we have just rewritten the soft-breaking terms in a hermitian basis. For large coefficients $a_i$, $b_i$, $c_i$ and $d_i$, the off-diagonal elements of the soft-breaking terms can still induce large flavour changes. MFV requires the coefficients to be of $\mathcal{O}(1)$, which then gives a non-trivial but suppressed off-diagonal flavour structure. Note that there has been no reduction of the number of free parameters. The hermiticity of $\mathbf{m}_L^2$ and $\mathbf{m}_E^2$ forces the coefficients $a_i$ and $b_i$ to be real, whereas there is no such constraint for $c_i$ and $d_i$. 

It will be important later on to summarise the CP-phases we have in our setup. Clearly, the spurion $\ynu^\dag \ynu$ introduces already several CP-violating phases: one Dirac phase, the two Majorana phases $\alpha$ and $\beta$ and the three seesaw parameters of the matrix $\Phi$. The second source of CP-violation are the MFV coefficients themselves. $a_i$, $b_i$, $c_i$ and $d_i$ contain all possible $G_f$ singlets, in particular traces of $\ye^\dag \ye$ and $\ynu^\dag \ynu$. But since the spurions carry complex phases, their traces can be complex and it would be inconsistent to require all MFV coefficients to be real. In fact, the $G_f$ symmetry does not protect the MFV coefficients from being complex and it is not even possible to define a CP-limit independently for the spurions and for the coefficients \cite{Mercolli2009}.

\section{Observables and CP-phases}

Equipped with this flavour structure of the soft-breaking terms, let us investigate the MSSM contributions to the leptonic EDMs and LFV processes. The one-loop contributions to the effective operator
\begin{equation}
 \mathcal{H}_{eff}=e\mathcal{M}^{IJ} \; \bar{\psi}_{L}^{I}\sigma_{\mu\nu}\psi_{R}^{J} \; F^{\mu \nu} \, + \, h.c.
\end{equation}
have already been calculated exactly \cite{Loop} and in the mass insertion approximation (MIA) \cite{Masina2002}. Only chargino and neutralino contributions will be considered, neglecting Barr-Zee \cite{BarrZee} and $\tan \beta$-enhanced non-holomorphic corrections \cite{Paradisi2008}.  The MIA calculation provides a good approximation even for light sparticles and it will help us to classify the different sources of CP-violation.

\begin{figure}
\centering
\psfrag{A}[t][Bb]{$\log \left[ \frac{\mbox{\tiny{BR}}(\mu \rightarrow e\gamma) }{\mbox{\tiny{BR}}_{exp}(\mu \rightarrow e\gamma)} \right]$}
\psfrag{B}[rc][Bb]{$ \log \left[  \frac{|d_e|}{|d_{e \; exp}|}  \right]$}
\psfrag{-}[r][Bb]{\small{-\phantom{|}}}
\psfrag{0}[c][Bc]{\small{0}}
\psfrag{2}[c][Bc]{\small{2}}
\psfrag{4}[r][Br]{\small{4}}
\psfrag{5}[r][BR]{\small{5}}
\psfrag{6}[r][Br]{\small{6}}
\psfrag{8}[r][Br]{\small{8}}
\psfrag{10}[r][Br]{\small{10}}
\psfrag{15}[r][Br]{\small{15}}
\psfrag{x}[r][Br]{\small{I}}
\psfrag{y}[r][Br]{\small{II}}
\psfrag{z}[r][Br]{\small{III}}
\includegraphics[bb=-57 -20 320 245,scale=0.8]{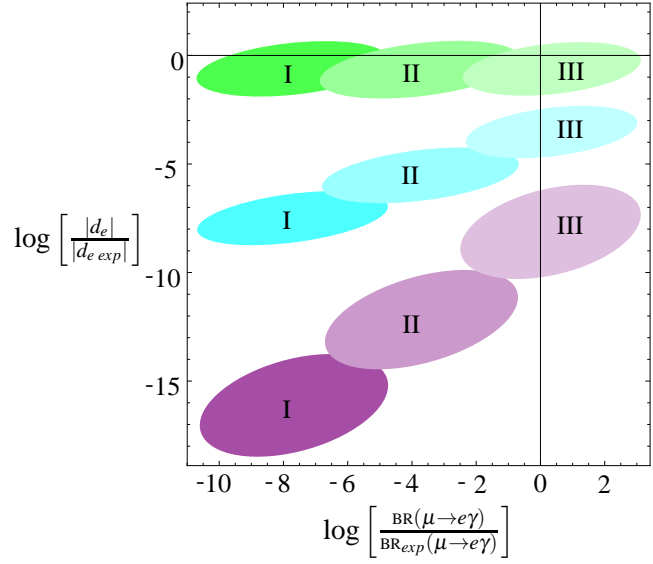} 
\caption{The $\mbox{BR}(\mu\rightarrow e\gamma) \, - \, d_e$ plane is plotted on a logarithmic scale, normalized to the experimental limits $\mbox{BR}(  \mu\rightarrow e\gamma)  <1.2\times10^{-11}$ and $|d_{e}|<1.6\;10^{-27}\;e\,$cm \cite{Exp}. The green contours represent the case for random MFV coefficients, whereas the turquoise ones have no flavour-blind phase, i.e. $c_1$ is real. The purple contours have all $c_i$ real, which is the case when only the flavour off-diagonal phases give contributions to the EDM. The neutrino parameters are varied (see \cite{Mercolli2009} for details) and the other parameters are fixed to: $\tan \beta = 10$, $m_{\nu1}=0$, $m_0=600$ GeV, $A_0=300$ Gev, $\mu=300$ Gev, $M_1=200$ GeV, $M_2=400$ GeV. The roman numbers distinguish different values of $M_R$: I: $ M_R =10^9$ GeV, II: $M_R =10^{11} $ GeV, and III: $ M_R =10^{13}$ GeV. }
\label{fig c1MR}
\end{figure}

We now randomly generate sets of $\mathcal{O}(1)$ MFV coefficients $a_i$, $b_i$, $c_i$, and $d_i$ and plot the resulting values of the observables in the $\mbox{BR}(\mu\rightarrow e\gamma) \, - \, d_e$ plane. In fig. \ref{fig c1MR}, 90\% contours are plotted, i.e. the area where 90\% of the calculated points lie. The different colours correspond to different settings of parameters.
In practice, not all the terms in the MFV expansion of eq. (\ref{eqn MFVexp}) give a significant contribution to the observables. To understand the main features of fig. \ref{fig c1MR}, it is sufficient to look at the dominant terms which are given by \cite{Mercolli2009}
\begin{equation} \label{eqn obs}
\begin{split}
& \mbox{BR}\left(  \mu \rightarrow e \, \gamma\right) \;  \sim  \; \mbox{BR}(\mu \rightarrow e \, \bar{\nu}_e \nu_\mu) \; \frac{M_{W}^{4} \, \alpha \, \tan^{2}\beta}{M_{SUSY}^4}  \left| \, \frac{a_{3}}{a_{1}} \left( \mathbf{Y}_{\nu}^{\dagger}\mathbf{Y}_{\nu} \right)^{12} \right|^{2} \;, \\[2ex]
& \frac{d_{e}}{e} \; \sim \; \frac{m_{e} \, \alpha }{M_{SUSY}^2}  \left(  \frac{\operatorname{Im}c_{1}}{a_{1}a_{7}%
}+\frac{\operatorname{Im}c_{3}\mathbf{Y}_{\nu}^{\dagger}\mathbf{Y}_{\nu}}%
{a_{1}a_{7}}-\frac{b_{1}\operatorname{Re}c_{3}}{a_{1}^{2}a_{7}}[\mathbf{Y}_{e}^{\dagger}\mathbf{Y}_{e},\mathbf{Y}_{\nu}^{\dagger}\mathbf{Y}_{\nu
}]\mathbf{Y}_{\nu}^{\dagger}\mathbf{Y}_{\nu} \right)  ^{11}\;.%
\end{split}
\end{equation}

The three cases I, II and III in fig. \ref{fig c1MR} show the dependence of the observables on the seesaw scale $M_R$. Since $\ynu^\dag \ynu$ is proportional to $M_R$, see eq. (\ref{eqn ynu}), $\mbox{BR}\left(  \mu \rightarrow e \, \gamma\right)$ is roughly proportional to $M^2_R$. Once the MSSM mass spectrum is fixed, such that the EDMs mostly respect their experimental constraint, we get a bound on $M_R$ from the LFV limit that does not allow for large values of $\ynu^\dag \ynu$.

In contrast to the LFV branching ratios, the EDMs are not always sensitive to a growing $M_R$. If the coefficient $c_1$ has a non-vanishing imaginary part (green contours), it will completely dominate and from eq. (\ref{eqn obs}) it is clear that the first term has no $M_R$ dependence. This dominance has been investigated already in ref. \cite{flavourblind} and could easily be explored by the future experiments. Setting $\mbox{Im} \, c_1 = 0$, the second term in eq. (\ref{eqn obs}) (turquoise contours) takes over and therefore also the EDM starts increasing with $M_R$. The same applies also to the case where only the third term gives a contribution to the EDM (purple contours), i.e. when $\mbox{Im} \, c_1 = \mbox{Im} \, c_3 = 0$. However, given the limit on $M_R$ from $\mu \rightarrow e \gamma$, these phases are never competitive with $\mbox{Im} \, c_1$.

We already mentioned the occurrence of several CP-violating phases in the previous section. The strong hierarchy between the green, turquoise and purple contours allows us to classify the various CP-phases according to their contributions to the leptonic EDMs \cite{Mercolli2009}.

\begin{itemize}
\item \emph{Flavour-blind phase.} Since all the $c_i$ and $d_i$ can be complex, the overall phase of the trilinear term $\mathbf{A}_e$ has to be defined depending on the convention for the phases of the flavour-blind sector (phase of $\mu$, etc.). In principle we can move this overall phase, i.e. the phase of  $c_1$, into the flavour-blind sector which would mean that MFV introduces only relative phases among the coefficients. However, once the MSSM phases are fixed and we allow $c_1$ to be complex, fig. \ref{fig c1MR} shows a striking dominance of the flavour-blind phase on the EDM of the electron. This follows clearly from eq. (\ref{eqn obs}) since the first term is not suppressed by the size of the spurions.

\item \emph{Flavour diagonal phases.} Still to leading order in the MIA, but already suppressed by the spurions, are the parameters $\mbox{Im} \, c_{i}$ for $i \neq 1$. The turquoise contours in fig. \ref{fig c1MR} show that these phases are suppressed by several orders of magnitude. In eq. (\ref{eqn obs}) the leading flavour diagonal phase $\mbox{Im} \, c_3 $ is shown in the second term. 

\item \emph{Flavour off-diagonal phases.} The phases of the spurions as well as $b_i$ and $d_i$ can only contribute at second order in the MIA. Since we have expanded the soft-breaking terms in a hermitian basis, the diagonal entries of the basis elements will always be real and therefore the flavour off-diagonal phases can only contribute when two of the soft-breaking terms are multiplied with each other. The strong suppression of these phases can be seen by means of the purple contours in fig. \ref{fig c1MR}. This scenario corresponds to dropping the first and second term in eq. (\ref{eqn obs}), where the third term is the dominant flavour off-diagonal contribution. 

\end{itemize}

This classification of CP-phases does not only apply to the lepton sector, since also the squark soft-breaking terms can be expanded in a similar MFV basis by replacing $\ye^\dag \ye$ with $\mathbf{Y}_d^\dag \mathbf{Y}_d$ and $\ynu^\dag \ynu$ with $\mathbf{Y}_u^\dag \mathbf{Y}_u$. A complete analysis of the new phases in hadronic CP-violating observables would be very desirable.

\section{Conclusion}

The MSSM, equipped with a seesaw mechanism of type I, is required to respect the principle of MFV. We have constructed the most general MFV expansion of the soft-breaking terms (eq. (\ref{eqn MFVexp})) and showed how new CP-violating phases appear. These phases are classified into flavour-blind, flavour diagonal and flavour off-diagonal, according to their strongly hierarchical contribution to the EDMs of the leptons. The numerical analysis shows a strong hierarchy between these classes, which can be understood from the structure of the MFV expansion and its connection to the MIA. 

Overall, we find that all the CP-violating phases allowed by the MFV principle are compatible with the experimental bounds in the lepton sector. Though a similar analysis has yet to be performed in the quark sector, the perspective for a rich CP-violating phenomenology within MFV looks very promising.

\section*{Acknowledgments}
I would like to thank Christopher Smith, Gilberto Colangelo and Stefan Lanz for the useful discussions and their help on the manuscript.
The Center for Research and Education in Fundamental Physics is supported by the \emph{Innovations- und Kooperationsprojekt C-13} of the Schweizerische Universit\"atskonferenz SUK/CRUS.

\end{document}